\documentclass[conference]{IEEEtran}
\IEEEoverridecommandlockouts

\usepackage[left=0.6in, right=0.6in, top=0.7in, bottom=0.6in]{geometry} 
\usepackage{cite}
\usepackage{amsmath,amssymb,amsfonts}
\usepackage{algorithmic}
\usepackage{algorithm}
\usepackage{graphicx}
\usepackage{array}
\usepackage{textcomp}
\usepackage{xcolor}
\def\BibTeX{{\rm B\kern-.05em{\sc i\kern-.025em b}\kern-.08em
    T\kern-.1667em\lower.7ex\hbox{E}\kern-.125emX}}

\begin{document}

\title{Multi-Tier UAV Edge Computing for Low Altitude Networks Towards Long-Term Energy Stability\\
}

\author{
\IEEEauthorblockN{Yufei Ye\textsuperscript{1}, Shijian Gao\textsuperscript{2}, Xinhu Zheng\textsuperscript{1,2}, and Liuqing Yang\textsuperscript{1,2}}
\IEEEauthorblockA{\textsuperscript{1}Intelligent Transportation Thrust, Hong Kong University of Science and Technology (Guangzhou), Guangzhou, China}
\IEEEauthorblockA{\textsuperscript{2}Internet of Things Thrust, Hong Kong University of Science and Technology (Guangzhou), Guangzhou, China}
\IEEEauthorblockA{Email: yye760@connect.hkust-gz.edu.cn, \{shijiangao, xinhuzheng, lqyang\}@hkust-gz.edu.cn}
}

\maketitle

\begin{abstract}
This paper presents a novel multi-tier UAV-assisted edge computing system designed for low-altitude networks. The system comprises vehicle users, lightweight Low-Tier UAVs (L-UAVs), and High-Tier UAV (H-UAV). L-UAVs function as small-scale edge servers positioned closer to vehicle users, while the H-UAV, equipped with more powerful server and larger-capacity battery, serves as mobile backup server to address the limitations in endurance and computing resources of L-UAVs. The primary objective is to minimize task execution delays while ensuring long-term energy stability for L-UAVs. To address this challenge, the problem is first decoupled into a series of deterministic problems for each time slot using Lyapunov optimization. The priorities of task delay and energy consumption for L-UAVs are adaptively adjusted based on real-time energy status. The optimization tasks include assignment of tasks, allocation of computing resources, and trajectory planning for both L-UAVs and H-UAV. Simulation results demonstrate that the proposed approach achieves a reduction of at least 26\% in transmission energy for L-UAVs and exhibits superior energy stability compared to existing benchmarks.
\end{abstract}

\section{Introduction}

Advancements in mobile edge computing (MEC) have greatly improved support for resource-intensive onboard applications in the Internet of Vehicles (IoV) \cite{yhou2024hierarchical}, such as environmental perception and high-precision map construction \cite{xcheng2022integrated, xcheng2022mmwave}. However, fixed terrestrial servers face limitations in service coverage and are prone to signal blockages, which hampers their adaptability to the dynamic demands of IoV. The rapid growth of the low-altitude economy (LAE) \cite{yjiang2025integrated} has positioned UAVs as key enablers, as their elevated positions and agile mobility provide better Line-of-Sight (LoS) channels, broader coverage, and enhanced scheduling capabilities \cite{gcheng2025networked}.

UAV-assisted edge computing in the IoV has recently attracted significant attention, leading to methods that optimize UAV positions, task offloading, and resource allocation. These aim to reduce vehicle energy consumption \cite{pzhao2025task, jwang2024anadaptive} and task execution delays \cite{myan2024edge, yliu2024mobile} while maximizing successful task completions \cite{zliao2024anadaptive}. However, UAVs’ compact designs limit their battery capacity and computing power, resulting in restricted endurance and increased risks of task deadline violations. Some studies have proposed high-altitude platforms (HAPs) as backup servers to address these limitations \cite{sli2024joint}, while increasing task success ratio \cite{ywang2024computation}, conserving energy of user devices \cite{ychen2024energy} or whole system \cite{hli2024dynamic}, and reducing sum of task delays and energy costs \cite{nwaqar2022computation}, but their altitude can lead to higher transmission energy and delays, further straining UAV energy reserves.

Prior research has also examined tiered UAV edge computing in low-altitude networks, typically involving a single UAV connected to a ground cloud server \cite{hyuan2024cost, bliu2023computation} or multiple upper-tier UAVs \cite{xren2024joint}. Most studies have focused on optimizing the trajectories of single-tier UAVs with fixed weights for task delays and UAV energy costs. However, dynamic delay-energy prioritization and joint optimization of multi-tier UAV trajectories are still underexplored.

In this paper, we propose an innovative multi-tier UAV-assisted edge computing system (MTUEC) designed for low-altitude networks, consisting of the vehicle users (task sources), the lightweight Low-Tier UAVs (L-UAVs), and a High-Tier UAV (H-UAV). The system aims to minimize the overall task execution delay while ensuring the long-term energy stability of L-UAVs. Positioned closer to vehicle users, L-UAVs function as lightweight aerial servers, handling a portion of the computational tasks. Given their limited computing power and energy reserves, an H-UAV equipped with a more powerful server and a larger-capacity battery operates at a higher altitude as a mobile compensatory server to support the L-UAVs by managing additional workloads.

MTUEC not only shortens the communication distance between the H-UAV and L-UAVs but also jointly optimizes their trajectories to improve the connectivity of inter-tier UAVs, both reducing L-UAV transmission energy and delays. The system employs Lyapunov optimization to adaptively balance task execution delays and L-UAV energy consumption based on their real-time energy states, ensuring their dynamic energy stability. Additionally, it jointly optimizes task assignment ratios, computing resource allocation, and flight trajectories for both L-UAVs and the H-UAV. The hierarchical architecture also facilitates workload balancing among UAVs, thereby enhancing the ecological stability of the entire system. Experimental results show that MTUEC achieves task delays comparable to or lower than existing benchmarks, reduces L-UAV transmission energy by at least 26\%, and maintains the most stable ratio of task delay to L-UAV energy deviation. This highlights the system's effectiveness in enhancing energy efficiency and responsiveness in dynamic network environments.


\section{System Model}

This section outlines the proposed multi-tier UAV edge computing system model, which encompasses the network model, communication model, and computation model.

\begin{figure}[htbp]
\centerline{\includegraphics[width=3.2 in]{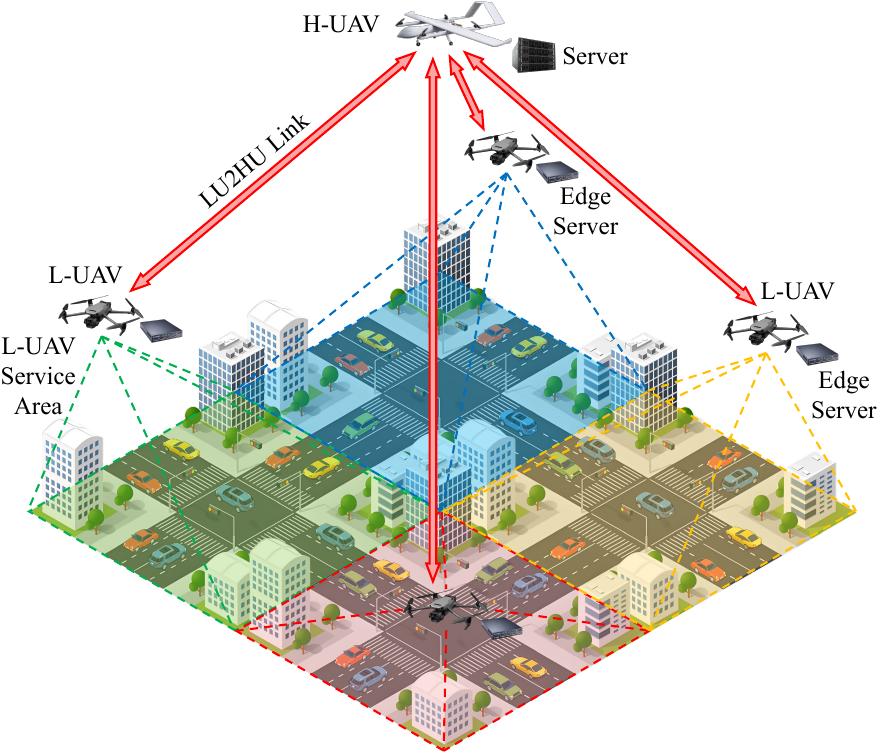}}
\caption{The proposed multi-tier UAV edge computing system.}
\label{fig_1}
\end{figure}

\vspace{-0.2cm}
\subsection{Network Model}

Fig.~\ref{fig_1} depicts the proposed system, comprising vehicle users in an urban traffic scenario, $U$ L-UAVs as edge servers with each having a designated service area, and one large-scale H-UAV as aerial backup server and central controller. Each service period is divided into $N$ time slots signified as $\mathcal{N}=\{0,1, ..., n, ...,N-1\}$ with length $\tau$ for each. We denote the set of vehicles in time slot $n$ and L-UAVs as $\mathcal{V}(n)=\{1,2, ..., v, ...,V(n)\}$ and $\mathcal{U}=\{1,2, ..., u, ...,U\}$, respectively. The horizontal positions of vehicle $v$, L-UAV $u$, and the H-UAV in time slot $n$ are signified by $\boldsymbol{G_v}(n) = (x_v(n), y_v(n))$, $\boldsymbol{G_u}(n) = (x_u(n), y_u(n))$, and $\boldsymbol{G_H}(n) = (x_H(n), y_H(n))$, with vertical coordinate of 0, $H_1$, and $H_2$, respectively \cite{jwang2024anadaptive, myan2024edge}. We also consider L-UAV flight energy consumption given by \cite{myan2024edge} $E_{u,flight}(n) = 0.5M_u \tau {\overline{s}_u(n)}^2$, where $M_u$ is the mass of L-UAV $u$, and $\overline{s}_u(n) = \frac{\|\boldsymbol{G_u}(n)-\boldsymbol{G_u}(n-1)\|}{\tau}$ is the average flight speed of L-UAV $u$ between time slot $n-1$ and $n$.

\subsection{Computation Model}

The task data size of vehicle $v$ in time slot $n$ is denoted as $D_v(n)$, with computational density $C_v(n)$ cycles/bit (i.e., number of CPU cycles required to process each bit) and a delay requirement $\tau_{v,max}(n)$. Each vehicle first transmits its initial task data to the associated L-UAV, which will compute a proportion $\alpha_v(n)$ of task. The remaining part is then transmitted to H-UAV for further computation. We signify the computing resource (i.e., CPU frequency) allocated by L-UAV $u$ to vehicle $v$'s task as $f_{v,u}(n)$. Thus, the L-UAV $u$'s computation delay and energy are $T_{v,u}^{comp}(n) = \frac{D_v(n)C_v(n)\alpha_v(n)}{f_{v,u}(n)}$ and $E_{v,u}^{comp}(n) = \kappa_u D_v(n)C_v(n)\alpha_v(n)f_{v,u}(n)^2$, respectively, where $\kappa_u$ denotes energy efficiency factor of L-UAV $u$’s processor \cite{jwang2024anadaptive}. The delay for H-UAV to compute vehicle $v$'s remaining task is $T_{v,H}^{comp}(n) = \frac{D_v(n)C_v(n)(1-\alpha_v(n))}{f_{v,H}(n)}$, where $f_{v,H}(n)$ denotes the computing resource allocated by H-UAV to this task.

\subsection{Communication Model}

Following \cite{pzhao2025task, myan2024edge}, the link between vehicle $v$ and L-UAV $u$ is modeled as Line-of-Sight (LoS) channel. The corresponding channel gain is $h_{v,u}(n)=\frac{\gamma_0}{d_{v,u}(n)^2}=\frac{\gamma_0}{H_1^2+\| \boldsymbol{G_u}(n) - \boldsymbol{G_v}(n)\|^2}$, where $d_{v,u}(n)$ is the distance between them at time slot $n$, $\gamma_0$ denotes the reference channel gain at the distance of 1m. We employ Orthogonal Frequency Division Multiple Access (OFDMA) technology \cite{ywang2024computation}, hence the transmission data rate between vehicle $v$ and L-UAV $u$ is given by $R_{v,u}(n) = B_{v,u}(n) \log_2 \left(1+\frac{P_v(n)h_{v,u}(n)}{N_0 B_{v,u}(n)}\right)$, where $B_{v,u}(n)$ denotes the allocated bandwidth for this channel, $P_v(n)$ is transmit power of vehicle $v$, and $N_0$ is noise power spectral density. The corresponding transmission delay is $T_{v,u}^{tr}(n) = \frac{D_v(n)}{R_{v,u}(n)}$. Similarly, the channel gain and data rate between L-UAV $u$ and H-UAV are $h_{u,H}(n) = \frac{\gamma_0}{(H_2-H_1)^2+\| \boldsymbol{G_H}(n) - \boldsymbol{G_u}(n)\|^2}$ and $R_{u,H}(n) = B_{u,H}(n) \log_2 \left(1+\frac{P_u(n)h_{u,H}(n)}{N_0 B_{u,H}(n)}\right)$, respectively, where $B_{u,H}(n)$ and $P_u(n)$ are the assigned bandwidth and transmit power of L-UAV $u$, respectively. The delay and energy incurred by L-UAV $u$ in transmitting the remaining task data of vehicle $v$ to H-UAV are $T_{v,u,H}^{tr}(n) = \frac{D_v(n)(1-\alpha_v(n))}{R_{u,H}(n)}$ and $E_{v,u,H}^{tr}(n) = P_u(n)\cdot T_{v,u,H}^{tr}(n)$, respectively.


\section{Problem Formulation and Transformation}

This section begins by formulating the primary optimization problem and then provides a detailed explanation of the problem transformation utilizing Lyapunov optimization.

\subsection{Problem Formulation}

Our objective is to minimize overall task delay $T(n) = \sum\nolimits_{v \in \mathcal{V}(n)} T_v(n)$ under the long-term L-UAV energy stability constraint through jointly optimizing task assignment ratio, computing resource allocation and trajectories of L-UAVs and H-UAV. The total execution delay for vehicle $v$'s task is $T_v(n) = T_{v,u}^{tr}(n) + T_{v,u}^{comp}(n) + T_{v,u,H}^{tr}(n) + T_{v,H}^{comp}(n)$. The energy consumed by L-UAV $u$ for vehicle $v$ is $E_{v,u}(n) = E_{v,u,H}^{tr}(n) + E_{v,u}^{comp}(n)$, hence the total energy consumption of L-UAV $u$ is $E_u(n) = \sum_{v \in \mathcal{V}_u(n)} E_{v,u}(n) + E_{u,flight}(n)$, where $\mathcal{V}_u(n)$ denotes set of vehicles in the service area of L-UAV $u$ in slot $n$. The prime optimization problem is formulated as

\textbf{Original problem:}
\vspace{-0.1cm}
\begin{align*}
\min _{\substack {\boldsymbol{\alpha}, \boldsymbol{f_U}, \boldsymbol{f_H}, \\ \boldsymbol{G_U}, \boldsymbol{G_H}}}
&\frac{1}{N}\sum_{n=0}^{N-1} \mathbb{E} [T(n)]\\
\text {s.t. } \hspace{0.2cm} &0\leq\alpha_{v}(n)\leq1,\forall v, n, \tag{1a}\\
& f_{v,u}(n)\geq0,f_{v,H}(n)\geq0,\forall v,u,n, \tag{1b}\\
& \sum \nolimits_{v \in \mathcal{V}_u(n)} f_{v,u}(n)\leq F_{u},\forall u,n, \tag{1c}\\
& \sum \nolimits_{v \in \mathcal{V}(n)}f_{v,H}(n)\leq F_H,\forall n,\tag{1d}\\
& T_{v}(n)\leq\tau_{v,max},\forall v,n, \tag{1e}\\
& \frac{1}{N} \sum_{n=0}^{N-1} \mathbb{E}[E_u(n)] \leq E_q, \forall u, \tag{1f}\\
& \frac{\|\boldsymbol{G_u}(n)-\boldsymbol{G_u}(n-1)\|}{\tau} \leq S_{u,max},\forall u,n,  \tag{1g}\\
& \frac{\|\boldsymbol{G_H}(n)-\boldsymbol{G_H}(n-1)\|}{\tau} \leq S_{H,max},\forall n.  \tag{1h}
\end{align*}

In terms of optimization variables, $\boldsymbol{\alpha} = \{ \alpha_v(n)| v\in\mathcal{V}(n), n\in\mathcal{N}\}$, $\boldsymbol{f_U} = \{ f_{v,u}(n)| v\in\mathcal{V}_u(n), u \in\mathcal{U}, n\in\mathcal{N}\}$, $\boldsymbol{f_H} = \{ f_{v,H}(n)| v\in\mathcal{V}(n), n\in\mathcal{N}\}$, $\boldsymbol{G_U} = \{ \boldsymbol{G_u}(n)| u \in\mathcal{U}, n\in\mathcal{N}\} $, and $\boldsymbol{G_H} = \{ \boldsymbol{G_H}(n)| n\in\mathcal{N}\} $. The constraint (1a) limits the range of task assignment ratios, while (1b) guarantees nonnegativity of allocated computing resources. (1c) and (1d) ensure the total allocated computing resource by each L-UAV and H-UAV does not exceed its available capacity, respectively. (1e) guarantees the delay requirement for each task is satisfied. In (1f), we define the reference energy quota per time slot of each L-UAV as $E_q = \frac{E_{full}}{N}$, where $E_{full}$ denotes L-UAV total energy when fully charged, hence it ensures the long-term energy stability of each L-UAV. (1g) and (1h) guarantee the flight speed of L-UAV $u$ and H-UAV will not exceed their maximum velocity $S_{u,max}$ and $S_{H,max}$, respectively.

\subsection{Problem Transformation}

In order to solve this problem, the overall system informations such as the number and position of vehicles, task data volume, and computation workloads for all time slots are required. However, the future system states are unknown at a specific time slot. To tackle this long-term optimization problem, we leverage the Lyapunov dynamic optimization to transform the original problem into a series of deterministic problems, which can be solved in an online manner in each time slot without the need for information of future system.

According to Lyapunov optimization, to minimize the overall task execution delay while ensuring energy stability of L-UAVs over long-term time span, we employ the drift-plus-penalty approach \cite{ychen2024energy}. As for the objective function of the original problem, the time-averaged task delay is suitable to be decoupled as a penalty term into the optimization target of each time slot. In order to decouple the long-term energy stability constraint (1f) of L-UAVs into each time slot, we establish energy quota deviation queues $\mathcal{Q}(n) = \{ Q_1(n), \ldots, Q_u(n), \ldots, Q_U(n)\}$ for L-UAVs. Specifically, the energy quota deviation queue of L-UAV $u$ at the beginning of time slot $n+1$ is expressed as
\begin{equation*}
Q_u(n+1) = \max\{ Q_u(n)+E_u(n)-E_q , 0 \}, \tag{2}
\end{equation*}
where $E_u(n)$ denote the energy consumption of L-UAV $u$ in the previous time slot $n$. Thus, it can indicate the cumulative deviation from energy quota of each L-UAV at the beginning of each time slot. The energy deviation queue at the initial time slot is $Q_u(0)=0, \forall u \in \mathcal{U}$. After removing constant terms in the objective function, the optimization problem within each time slot is transformed into

\textbf{Transformed problem:}
\vspace{-0.2cm}
\begin{align*}
\min _{\substack {\boldsymbol{\alpha}, \boldsymbol{f_U}, \boldsymbol{f_H}, \\ \boldsymbol{G_U}, \boldsymbol{G_H}}}
& K \cdot T(n)+\sum_{u=1}^UQ_u(n) \cdot E_u(n)\\
\text {s.t. } \hspace{0.2cm} & \text{(1a)-(1e), (1g), and (1h)},
\end{align*}
where $K>0$ is Lyapunov control parameter. The objective function is a dynamic weighted sum of total task delay and energy consumption of L-UAVs, with $Q_u(n)$ of each L-UAV $u$ serving as variable weight for its energy consumption. Consequently, the L-UAV with a larger $Q_u(n)$ implies a greater deviation of its energy consumption from quota at the beginning of the current time slot, thus increasing the effort to conserve its energy. Conversely, with small or zero $Q_u(n)$, we put greater emphasis on reducing task execution delay. In this way, we achieve the goal of dynamically adjusting the tradeoff between task execution delay and energy consumed by L-UAVs based on their real-time energy status.


\section{Joint Task Assignment, Resource Allocation, and UAV Trajectory Optimization}

In this section, we elaborate on the problem decomposition and the solving process for each subproblem, and finally summarize the overall algorithm. The transformed problem is a non-convex optimization problem involving five categories of interdependent optimization variables. We can elegantly decompose it into several low-dimensional and tractable subproblems, thereby reducing the solving complexity. This naturally motivates the adoption of the block coordinate descent (BCD) method, which is well-suited for large-scale optimization.


\subsection{Joint Optimization of Task Assignment Ratio and Computing Resource Allocation}

The core idea of the BCD method is to decompose the complex problem into multiple subproblems by reasonably grouping the optimization variables, which enables us to design suitable solution tailored to their distinct structures. Each subproblem is then iteratively optimized to ultimately converge into an efficient solution. Based on this fundamental idea, we decompose the transformed problem into three subproblems: joint optimization of task assignment ratios and computing resource allocation for L-UAVs and H-UAV, trajectory optimization of L-UAVs and that of H-UAV. Specifically, the ratio of each vehicle's task assigned to the two types of UAVs and the corresponding computing resource allocated by them are strongly correlated and exhibit high sensitivity to the numerical changes of each other. Therefore, to reduce the number of overall iterations, thereby improving the convergence efficiency, we group these three categories of variables together for joint optimization. With the trajectories of L-UAVs and H-UAV are presumed to be fixed and retaining only related terms in the objective function and constraints, this subproblem can be formulated as

\textbf{Subproblem-1: Joint optimization of task assignment ratios and computing resource allocation for both L-UAVs and H-UAV}
\vspace{-0.1cm}
\begin{align*}
\min _{\substack {\boldsymbol{\alpha}, \boldsymbol{f_U}, \boldsymbol{f_H}}}
&K \cdot \sum_{v=1}^{V(n)} T_v^{comp}(n) + \sum_{u=1}^U Q_u(n) \cdot E_u(n)\\
\text {s.t. } \hspace{0.2cm} & \text{(1a)-(1e)},
\end{align*}
where $T_v^{comp}(n)=T_{v,u}^{comp}(n) + T_{v,u,H}^{tr}(n) + T_{v,H}^{comp}(n)$. It can be proven to be a convex optimization problem with determined $\alpha_v (n)$, which can be optimally solved by standard convex optimization tools such as CVX solver. When $f_{v,u}(n)$ and $f_{v,H}(n)$ are given, the objective function is linear with $\alpha_v (n)$, hence the optimal value can be found by linear search method. Based on the above analysis, we utilize an alternating approach to solve subproblem-1. Specifically, we initially assign a feasible value to each $\alpha_v (n)$, based on which we optimize $f_{v,u}(n)$ and $f_{v,H}(n)$. Subsequently, we update $\alpha_v (n)$ using the once-optimized $f_{v,u}(n)$ and $f_{v,H}(n)$. We repeat those steps until convergence, thereby obtaining the final solution.

\subsection{L-UAV Trajectory Optimization}

Given the task assignment ratios and computing resource allocation strategies, we then optimize the trajectories for L-UAVs and H-UAV. The flight trajectories of L-UAVs are tightly coupled with vehicle to L-UAV (V2LU) communications and their flight energy. We optimize their trajectories in order to balance their communications with multiple vehicle users they serve and their real-time energy status through minimizing the adaptive weighted sum of overall V2LU data transmission delay and L-UAV flight energy under the task delay requirement and L-UAV maximum velocity constraints. The challenge lies in the fact that it is a non-convex problem since the objective function and constraint (1e) are non-convex with $\boldsymbol{G_u}(n)$. The objective function to be minimized and the left-hand side of constraint (1e), which must be kept below a certain threshold, are well-suited for relaxation by their convex upper bounds. Based on this insight, we adopt the successive convex approximation (SCA) technique to first transform the original problem into a sequence of approximated convex problems and then continuously solve them until convergence.

In the objective function, $T_{v,u}^{tr}(n)$ is the non-convex term. Its denominator $R_{v,u}(n)$ is also a non-convex function of $\boldsymbol{G_u}(n)$, but it is convex w.r.t. $\| \boldsymbol{G_u}(n) - \boldsymbol{G_v}(n)\|^2$. We define $\varphi_{v,u} (n) = \| \boldsymbol{G_u}(n) - \boldsymbol{G_v}(n)\|^2$, hence $R_{v,u}(n)$ can be globally lower-bounded by its first-order Tayler expansion with $\varphi_{v,u} (n)$ at any point. We denote $\boldsymbol{G_u^k}(n)$ as the position of L-UAV $u$ at the $k$-th iteration. Then the lower bound of $R_{v,u}(n)$ can be derived as
\vspace{-0.2cm}
\begin{equation*}
\widehat{R}_{v,u}(n)=R_{v,u}^k(n)+\nabla R_{v,u}^k(n)\left(\varphi_{v,u}(n)-\varphi_{v,u}^k(n)\right), \tag{3}
\end{equation*}
where $R_{v,u}^k(n)$ and $\nabla R_{v,u}^k(n)$ signify the data transmission rate between vehicle $v$ and L-UAV $u$ and the first-order derivative of $R_{v,u}(n)$ w.r.t. $\varphi_{v,u} (n)$ at the $k$-th iteration, respectively. Specifically, they are given by
\vspace{-0.05cm}
\begin{equation*}
R_{v,u}^k(n) = B_{v,u}(n) \log_2 \left(1+\frac{P_v(n)\delta}{ H_1^2 + 
\varphi_{v,u}^k(n) }\right), \tag{4}
\end{equation*}
\vspace{-0.05cm}
\begin{equation*}
\nabla R_{v,u}^k(n) = \frac{- B_{v,u}(n)P_v(n) \delta \log_2e}{(H_1^2 + \varphi_{v,u}^k(n)) (H_1^2 + \varphi_{v,u}^k(n) + P_v(n) \delta)}, \tag{5}
\end{equation*}
where $\varphi_{v,u}^k(n) = \| \boldsymbol{G_u^k}(n) - \boldsymbol{G_v}(n) \|^2$ and $\delta = \gamma_0 / N_0$. Consequently, we obtain a convex upper bound for the non-convex term in the objective function, which is given by $T_{v,u}^{tr}(n) \leq \frac{D_v(n)}{\widehat{R}_{v,u}(n)}$. Since $T_{v,u}^{tr}(n)$ is also contained in constraint (1e), following a similar derivation, a convex upper bound for the left-hand side of (1e) can be expressed as $T_v(n) \leq \frac{D_v(n)}{\widehat{R}_{v,u}(n)} + T_{v}^{comp}(n)$.

Through the above transformation, we convert the original non-convex problem into its convex approximation problem. Hence the subproblem of L-UAV trajectory optimization is reformulated as

\textbf{Subproblem-2: L-UAV trajectory optimization}
\vspace{-0.1cm}
\begin{align*}
\min _{\substack {\boldsymbol{G_U}(n)}}
& K \cdot \sum_{u=1}^U \sum_{v=1}^{V_u(n)} \frac{D_v(n)}{\widehat{R}_{v,u}(n)} + \sum_{u=1}^U Q_u(n) \cdot E_{u,flight}(n)\\
\text {s.t. } \hspace{0.2cm} & \text{(1g),} \\
& \frac{D_v(n)}{\widehat{R}_{v,u}(n)} + T_{v}^{comp}(n) \leq \tau_{v,max}, \forall v,
\end{align*}
which can be solved by CVX solver iteratively. The subproblem in each iteration is formed based on the solution from previous iteration. Through iteratively solving these subproblems, it ultimately converges to an efficient near-optimal solution of the original problem that satisfies all constraints.

\subsection{H-UAV Trajectory Optimization}

Given the trajectories of L-UAVs, task assignment ratios and resource allocation strategy, we further optimize H-UAV's trajectory to facilitate its collaboration with multiple L-UAVs based on their path design. H-UAV trajectory optimization holds dual significance: (1) reducing the data transmission delay from L-UAVs to H-UAV (LU2HU) and (2) conserving the transmission energy for each L-UAV according to its real-time energy condition $Q_u(n)$. To this end, we should minimize the dynamic weighted sum of LU2HU data transmission delay and transmit energy consumption of each L-UAV, subject to delay requirement constraint and H-UAV maximum speed limitation.

The terms $T_{v,u,H}^{tr}(n)$ and $E_{v,u,H}^{tr}(n)$ are non-convex in the objective function w.r.t. $\boldsymbol{G_H}(n)$. Fortunately, the non-convexity of the latter stems exclusively from its dependence on the former. Thus, we only need to deal with $T_{v,u,H}^{tr}(n)$. Similar to the aforementioned analysis, since $R_{u,H}$ is convex with $\| \boldsymbol{G_H}(n) - \boldsymbol{G_u}(n)\|^2$, we define $\varphi_{u,H}(n) = \| \boldsymbol{G_H}(n) - \boldsymbol{G_u}(n)\|^2$. Hence the convex upper bound of the non-convex term $T_{v,u,H}^{tr}(n)$ in objective function can be derived as $T_{v,u,H}^{tr}(n) \leq \frac{D_v(n) ( 1- \alpha_v(n) )}{\widehat{R}_{u,H}(n)}$, where $\widehat{R}_{u,H}(n)$ is given by
\begin{equation*}
\widehat{R}_{u,H}(n) = R_{u,H}^i(n) + \nabla R_{u,H}^i(n)\left(\varphi_{u,H}(n)-\varphi_{u,H}^i(n)\right). \tag{6}
\end{equation*}
In (6), $R_{u,H}^i(n)$ and $\nabla R_{u,H}^i(n)$ denote the data transmission rate from L-UAV $u$ to H-UAV and the first-order derivative of $R_{u,H}(n)$ w.r.t. $\varphi_{u,H}(n)$ at the $i$-th iteration, respectively. Due to space limitation, we omit their specific expressions. Consequently, the subproblem of H-UAV trajectory optimization is converted into the convex approximation problem as

\textbf{Subproblem-3: H-UAV trajectory optimization}
\begin{align*}
\min _{\substack {\boldsymbol{G_H}(n)}}
& \sum_{u=1}^U \left( K + Q_u(n) P_u(n) \right) \sum_{v=1}^{V_u(n)} \frac{D_v(n) ( 1- \alpha_v(n) )}{\widehat{R}_{u,H}(n)}\\
\text {s.t. } \hspace{0.2cm} & \text{(1h)}, \\
& T_{v,u}^{tr}(n) + T_{v,u}^{comp}(n) + \frac{D_v(n) ( 1- \alpha_v(n) )}{\widehat{R}_{u,H}(n)} \\
& + T_{v,H}^{comp} \leq \tau_{v,max}, \forall v,
\end{align*}
which can be iteratively solved by CVX solver. By this point, we complete the trajectory optimization for L-UAVs and H-UAV.

The steps of the overall algorithm in each time slot $n$ are outlined in the \textbf{Algorithm 1}. We first initialize an array of system parameters including the vehicle and L-UAV sets, data volume and computational intensity of each task, computing capacities and current positions of L-UAVs and H-UAV, followed by updating energy quota deviation queue for each L-UAV. Since the joint optimization of task assignment and computing resource allocation depends on the trajectory optimization results of the two types of UAVs from the previous iteration and H-UAV trajectory optimization relies on the L-UAV trajectories optimized in the current iteration, we then iteratively solve Subproblem-1, Subproblem-2, and Subproblem-3 in sequence, while holding the remaining variables fixed at their respective most recently optimized values, until convergence or the maximum number of iterations is reached. The superscript $j$ of each variable represents the current solution at the beginning of the $j$-th iteration.

\begin{algorithm}[!t]
    \caption{Joint Optimization Algorithm of Task Assignment, Computing Resource Allocation, and UAV Trajectories.}
    \label{alg:overall}
    \renewcommand{\algorithmicrequire}{\textbf{Input:}}
    \renewcommand{\algorithmicensure}{\textbf{Output:}}
    
    \begin{algorithmic}[1]
        \REQUIRE Vehicle set $\mathcal{V}(n)$, L-UAV set $\mathcal{U}$, task profile $D_v(n)$ and $C_v(n)$, L-UAV computing capacities $F_u$, H-UAV computing capacity $F_H$, vehicle positions $\boldsymbol{G_v}(n)$, L-UAV positions $\boldsymbol{G_u}(n-1)$, and H-UAV position $\boldsymbol{G_H}(n-1)$, $\forall v,u$.   
        \ENSURE Task assignment ratios $\boldsymbol{\alpha}(n)$, L-UAV and H-UAV computing resource allocation $\boldsymbol{f_U}(n)$ and $\boldsymbol{f_H}(n)$, L-UAV and H-UAV trajectories $\boldsymbol{G_U}(n)$ and $\boldsymbol{G_H}(n)$, respectively.     
        
        \STATE Update $Q_u(n)$ according to (2) or set $Q_u(n) = 0$ at the initial time slot, and initialize iteration number $j=0$.

        \REPEAT
            \STATE Solve \textbf{Subproblem-1} to obtain $\boldsymbol{\alpha}^{j+1}, \boldsymbol{f_U}^{j+1}, \boldsymbol{f_H}^{j+1}$ with given $\boldsymbol{G_U}^j$ and $\boldsymbol{G_H}^j$.
            \STATE Solve \textbf{Subproblem-2} to obtain $\boldsymbol{G_U}^{j+1}$ with given $\boldsymbol{\alpha}^{j+1}, \boldsymbol{f_U}^{j+1}, \boldsymbol{f_H}^{j+1}$, and $\boldsymbol{G_H}^j$.
            \STATE Solve \textbf{Subproblem-3} to obtain $\boldsymbol{G_H}^{j+1}$ with given $\boldsymbol{\alpha}^{j+1}, \boldsymbol{f_U}^{j+1}, \boldsymbol{f_H}^{j+1}$, and $\boldsymbol{G_U}^{j+1}$.
            \STATE Update objective function value based on above variables.
            \STATE Update $j=j+1$.

        \UNTIL The objective value converges or $j>j_{max}$.

    \end{algorithmic}
\end{algorithm}


\section{Simulation Results}

In this section, we conduct simulation experiments to evaluate the performance of the proposed method. We consider a 1 km$\times$1 km square traffic area, where vehicles are randomly distributed, with the speed ranging from 30 to 80 km/h \cite{nwaqar2022computation}. We deploy four L-UAVs and one H-UAV server with maximum velocity of 15 m/s \cite{myan2024edge}. To ensure uniform spatial distribution, the initial positions of four L-UAVs are set at (250, 250), (750, 250), (750, 750), and (250, 750) with each serving the area of 500 m$\times$500 m centered at its respective location, while the H-UAV is initially positioned at the center of entire area. Following \cite{myan2024edge, sli2024joint, hyuan2024cost}, the parameter setup is summarized in Table~\ref{tab:table1}.

\vspace{-0.4cm}
\begin{table}[!h]
\caption{Simulation Parameters}
\label{tab:table1}
\centering
\begin{tabular}{|m{1.5cm}<{\centering}|m{2.0cm}<{\centering}|m{1.5cm}<{\centering}|m{2.0cm}<{\centering}|}
\hline
\textbf{Parameter} & \textbf{Value} & \textbf{Parameter} & \textbf{Value}\\
\hline
$V$ & 10 $\sim$ 40 & $N_0$ & -174 dBm/Hz\\
\hline
$D_v$ & 1$\sim$10 Mb & $B_{v,u}$ & 2 MHz\\
\hline
$C_v$ & 10 $\sim$ 100 & $B_{u,H}$ & 10 MHz\\
\hline
$\tau_{v,max}$ & 50$\sim$200 ms & $E_q$ & 4.5 J\\
\hline
$F_u$ & 10 GHz & $k_u$ & $10^{-27}$\\
\hline
$F_H$ & 50 GHz & $M_u$ & 4 Kg\\
\hline
$P_v$ & 0.5 W & $H_1$ & 100 m\\
\hline
$P_u$ & 1 W & $H_2$ & 150 m\\
\hline
$\gamma_0$ & -50 dB & $\tau$ & 0.2 s\\
\hline

\end{tabular}
\end{table}

We compare the performance of the proposed MTUEC method with the following benchmark methods:

\begin{itemize}
    \item \textbf{MTUEC with Fixed H-UAV Trajectory (FT-MTUEC)}: The H-UAV in the proposed method is modified to follow a fixed trajectory traversing along the diagonal of the field.
    \item \textbf{HAP-UAV Collaborative Resource Allocation Algorithm (HURA)}: The HAP-UAV collaborative resource optimization algorithm in \cite{sli2024joint} minimizing task delay with fixed energy constraint in each time slot.
    \item \textbf{UAV-Terrestrial Server Collaborative Delay-Centric Algorithm (UTDC)}: The UAV-terrestrial server collaborative task delay minimization method in \cite{yliu2024latency} without energy limitation.
\end{itemize}

\begin{figure}[!t]
\centerline{\includegraphics[width=2.5 in]{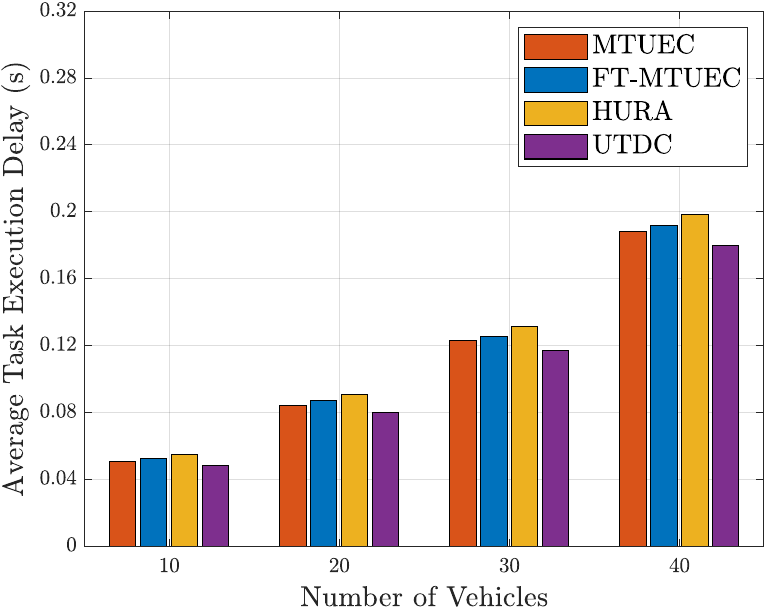}}
\caption{The comparison of average task execution delay among different methods with varying
numbers of vehicles.}
\label{fig_2}
\end{figure}

Fig.~\ref{fig_2} illustrates the average task execution delay for different methods with varying numbers of vehicles. Given that merely minimizing task delay is not the dominant strength of MTUEC, Fig.~\ref{fig_2} indicates that it can achieve comparable or even lower time-average task delay than benchmarks on the basis of flexibly trading off between delay and L-UAV energy consumption. Compared to HURA with fixed energy limit in each time slot, MTUEC will fully commit to minimizing task delay when there is no L-UAV energy deviation, thereby achieving around 8\% reduction in time-average delay. Since UTDC solely prioritizes delay minimization, its average delay is inevitably lower than that of MTUEC, yet it is only reduced by up to 5\%. The unoptimized H-UAV trajectory in FT-MTUEC increases the LU2HU transmission delay and energy, resulting in roughly 3\% increment of total task delay compared with MTUEC.

\begin{figure}[!t]
\centerline{\includegraphics[width=2.5 in]{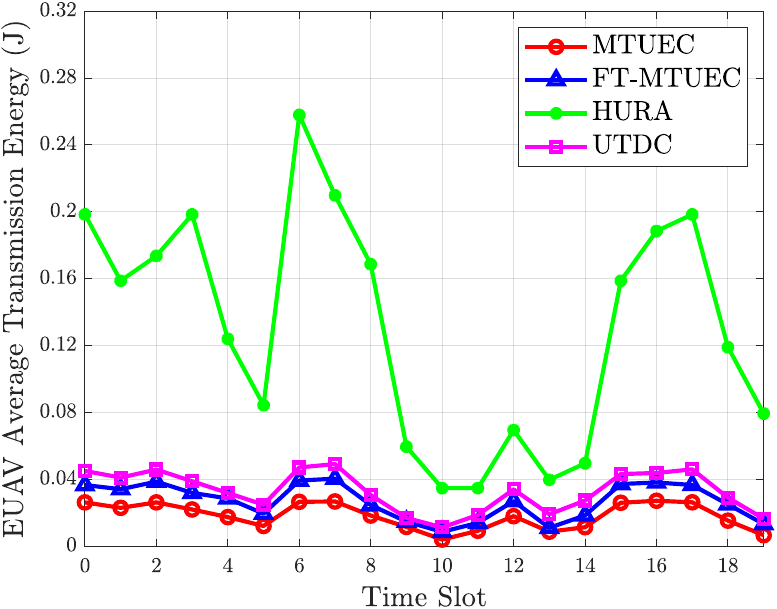}}
\caption{The comparison of average L-UAV transmission energy among different methods.}
\vspace{-0.4cm}
\label{fig_3}
\end{figure}

Fig.~\ref{fig_3} shows the average L-UAV transmission energy of these four methods over time slots. Benefiting from the short communication distance between L-UAVs and H-UAV as well as the optimized trajectory of H-UAV, MTUEC achieves the minimum L-UAV transmission energy. It attains 26\% reduction on average compared to FT-MTUEC, since the fixed H-UAV trajectory in FT-MTUEC prevents H-UAV from balancing communications with multiple L-UAVs. This highlights the necessity of H-UAV trajectory optimization. Due to the excessive height of HAP and the impractical trajectory design resulting from its cumbersome size, HURA incurs the highest L-UAV transmission energy, which is 3 to 9 times greater than that of MTUEC. UTDC employs ground base station as backup server with fixed location and greater distance to L-UAVs than H-UAV, which makes MTUEC reduce the L-UAV transmission energy by an average of 41\% compared to UTDC. These further underscore the advantage of low-altitude tiered UAV architecture.

\begin{figure}[!t]
\centerline{\includegraphics[width=2.5 in]{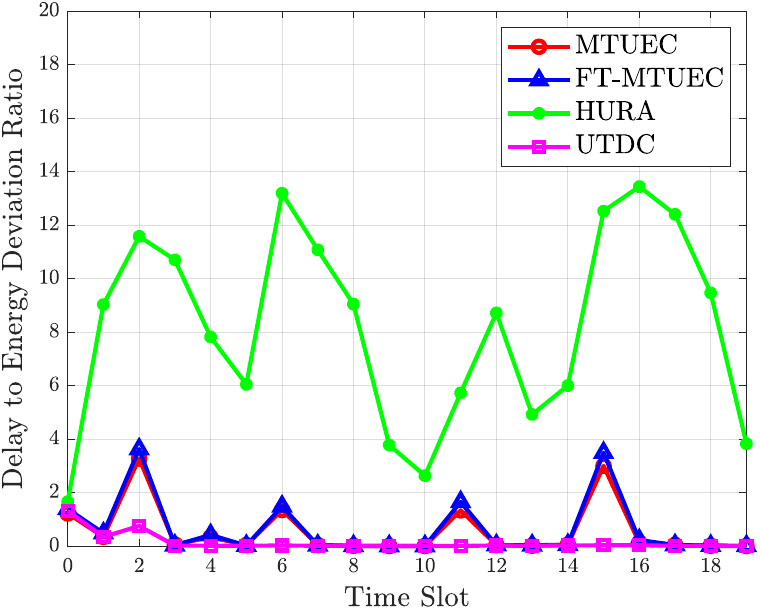}}
\caption{The comparison of the delay to energy deviation ratio among different methods.}
\vspace{-0.4cm}
\label{fig_4}
\end{figure}

To reflect the advantage of MTUEC in flexibly trading off between task execution delay and L-UAV energy consumption, we define the Delay to Energy Deviation Ratio (DEDR) as the ratio of the average task delay to the average energy quota deviation of L-UAVs. A small constant is added to denominator to prevent division by zero when there is no energy deviation. An increase in L-UAV energy deviation indicates a higher degree of its energy shortage in the current time slot. Consequently, the MTUEC will trade task delay for conserving L-UAV energy consumption, and vice versa. Therefore, a stabler DEDR implies a better capability in delay-energy tradeoff. Fig.~\ref{fig_4} illustrates that MTUEC exhibits the most stable DEDR across time slots. FT-MTUEC undergoes a slightly larger fluctuation, with an average increase of 15\%, since the trajectory of H-UAV is not optimized, which results in marginally reduced system stability. The fluctuation of HURA is much more pronounced, 14 times higher than MTUEC on average, because when the task volume surges, the task delay also significantly increases due to the strict single-slot energy constraint of L-UAVs, and vice versa. Since UTDC focuses solely on minimizing task delay without energy constraint, it leads to a continuous accumulation of L-UAV energy deviation. The progressively increasing denominator causes its DEDR to gradually approach zero.


\section{Conclusion}
This paper proposed a multi-tier UAV edge computing system for low-altitude networks, aiming to minimize system task delay while maintaining the energy stability of L-UAVs. The long-term optimization problem was decoupled into a series of problems solved online for each time slot using Lyapunov optimization, allowing for a dynamic tradeoff between task execution delay and L-UAV energy consumption based on their instantaneous energy states. We jointly optimized task assignment ratios, computing resource allocation, and trajectories for both the L-UAVs and the H-UAV. Simulation results validated that the proposed method outperformed benchmarks in maintaining relative stability between task delay and L-UAV energy quota deviation while achieving comparable or lower time-average delay.



\end{document}